# Laser boron fusion reactor with picosecond petawatt block ignition

Heinrich Hora[1]*, Shalom Eliezer[2], Jiaxiang Wang[3], Georg Korn[4], Noaz Nissim[2], Yanxia Xu[3], Paraskevas Lalousis[5], Götz Kirchhoff[6] and George H. Miley[7]

[1]Department of Theoretical Physics, University of New South Wales, Sydney 2052, Australia
[2]SOREQ Research Centre, Yavne, Israel
[3]State Key Laboratory of Precision Spectroscopy, East China Normal University, Shanghai 200062, China
[4]ELI-Beam, Prague, Czech Republic
[5]Institute of Electronic Structure and Laser FORTH, Heraklion, Greece
[6]UJG Management GmbH, 85586 Poing, Germany
[7]Dept. Nuclear, Plasma & Radiol. Engineering, University of Illinois, Urbana IL, USA
*h.hora@unsw.edu.au

Abstract. Fusion of hydrogen with the boron Isotope 11, HB11 at local thermal equilibrium LTE, is $10^5$ times more difficult than fusion of deuterium and tritium, DT. If - in contrast - extreme non-equilibrium plasma conditions are used with picoseconds laser pulses of more than 10PW power, the difficulties for fusion of HB11 change to the level of DT. This is based on a non-thermal transfer of laser energy into macroscopic plasma motion by nonlinear (ponderomotive) forces as theoretically predicted and experimentally confirmed as "ultrahigh acceleration". Including elastic nuclear collisions of the alpha particles from HB11 reactions results in an avalanche process such that the energy gains from HB11 fusion is nine orders of magnitudes above the classical values. In contrast to preceding laser fusion with spherical compression of the fuel, the side-on direct drive fusion of cylindrical uncompressed solid boron fuel trapped by magnetic fields above kilotesla, permits a reactor design with only one single laser beam for ignition within a spherical reactor. It appears to be potentially possible with present day technology to build a reactor for environmentally fully clean, low-cost and lasting power generation.

I. INTRODUCTION

The generation of energy from nuclear fusion reactions of hydrogen to heavier elements is the source of radiation emission as known from nearly all observed stars in the Universe. To use this in a controlled way of a power station on earth is an important aim where two basically different ways are followed up. One is aiming a continuously burning of high temperature plasma by using magnetic confinement of low density plasma in the necessary very large size thermonuclear experimental reactor ITER [1], and the other is the pulsating energy source with miniaturization of Edward Teller's first manmade energy generating fusion explosions [2] to the very small size [3] for a controlled operation in a power reactor. The discovery of the laser for ignition in such pulsating reactions was motivated as solution from the beginning. This way has been demonstrated by experiments using intense x-rays instead of lasers to produce ignitions with high energy gains [4]. The possible most sophisticated laser properties are expected to lower the needed ignition thresholds of the radiation.

Laser fusion of the easiest reaction uses heavy and superheavy hydrogen deuterium D and tritium T. It has reached the highest gains with the biggest laser on earth at the NIF facility in Livermore/California. The gains are not very far below breakeven. When these developments will come closer to the consideration for designing a power station, there will be two problems to be solved. One question is the handling of nuclear radiation with generation and processing of tritium including waste generation by the neutrons from the DT reaction. Their number is about four times higher per generated energy than the neutron production in uranium fission reactors as it is a well known problem also for the ITER reactor [5], with reference to the views of Sir William Mitchell. The second question is how to operate a close sequence of the reactions for the pulsating energy generation at radial laser irradiation for the condition of DT and how simultaneously the generated fusion energy can be recuperated from within the apparatus.

The following considerations for the pulsating reactor are about eventual possibilities how the mentioned two difficulties from nuclear radiation and of the spherical irradiation geometry of the laser irradiation may be eliminated by an option of laser fusion using a new approach for the long known fusion of hydrogen H reacting with the boron isotope 11 (HB11 fusion). This laser boron fusion is based on the ultrahigh acceleration of plasma blocks at such extremely high laser intensities that the non-thermal conversion of laser energy into macroscopic motion of plasma blocks is dominating. The combination with laser generated very high magnetic fields is needed for trapping a cylindrical plasma volume. Furthermore the HB11-reaction is profiting from the now measured unique type of an avalanche reaction.

II. ANEUTRONIC CLEAN BORON FUSION

The measurement of the neutron-free nuclear fusion reaction of light hydrogen H with the boron isotope 11 goes back to the measurements of Oliphant and Lord Rutherford [6] by irradiation of protons of less than MeV energy on boron targets

$$H + {}^{11}B = 3\,{}^{4}He + 8.7\,\text{MeV} \qquad (1)$$

Where the generated energy is distributed to equal parts to the generated helium nuclei as alpha particles. This highly desired aneutronic reaction cannot be used for a power station under the plasma conditions of local thermal equilibrium LTE at the conditions of continuously working devices as the ITER because of too high energy losses by cyclotron radiation. The laser irradiation on spherical HB11 fuel for ablation force compression, heating and thermonuclear ignition has been studied extensively (see Chapter 9.6 of [7]) and an exothermal energy gain need an ignition density above 100.000 times the solid state. Comparing this with the highest calculated direct drive volume ignition of DT fusion, the difficulty against DT is altogether given by a similar *factor of fife orders of magnitude*, even if the latest measurement of fusion reaction cross sections with resonances [8] were used. This is the reason that the laser-fusion experts considered the HB11 fusion as impossible.

The advantage of a neutron-free fusion of HB11 was indeed so highly attractive [9][10], that numerous attempts were considered which all had to be based on plasma conditions with non-thermal equilibrium [11] as a necessary condition. These were given as Hirsch-Miley [12][13] "electrostatic confinement" following a proposal by Farnsworth and later work [14][15] or the modification as inverse magnetic field conFiguration [16], or in the ion beam dominated MIGMA [17] or plasma focus conditions [18]. These low plasma density conFigurations in the category of ITER have not led to a measurement of HB11 reactions [19]. Only at high density conditions with plasmas in the range up to the solid state, reactions were measured by irradiation of laser pulses on uncompressed targets resulting per laser shot in 1000 reactions [20], in more than one million [21] and in a billion [22] with detailed repetition and some increase [23].

Computations for non-thermal equilibrium conditions of laser ignition of solid state fusion targets were performed [24][25] with the result that for DT fusion, an irradiated energy density within one picosecond had to have an energy flux threshold in the range of few times of R = $10^8$ J/cm$^2$ had to initiate the fusion flame for the reaction in the uncompressed solid density fusion fuel the for plane wave front and target geometry. These computations had to be updated in view of collective collisions and inhibition of thermal conduction due to double layer effects [26][27]. When instead of DT the fusion reaction cross section of HB11 were used [28][29], the energy flux threshold R arrived in the same range R. This was the surprising result for HB11 compared with DT that the ignition had an *increase of the fusion condition fife orders of magnitudes higher against the classical value* at thermal equilibrium.

III. PIC COMPUTATION OF THE PLASMA BLOCK ACCELERATION FOR THE FUSION FLAME

The essential mechanisms for the generation of the fusion flame is the deposition of the energy flux in the range within one picoseconds into the solid fuel surface. This was achieved on the basis of the ultrahigh acceleration of plasma bock by dominating nonlinear forces, e.g. when a $10^{18}$ W/cm$^2$ laser pulse was hitting a nearly critical density deuterium block with an initially low reflection density profile. After one nanosecond interaction, a plasma block had been accelerated with $10^{20}$ cm/s$^2$ moving against the laser beam and another block on the other side moving into the target, Fig. 1, (see Figures 10.18a and b of [30] of Fig. 8.4 of [7]) in exact agreement with later measurements [31][32]. This was not a simple radiation pressure process [32a] but a dielectric explosion of both the blocks [33][34][35] due to the decreasing of the dielectric constant of the plasma causing an increase of the electromagnetic energy density of the laser field [7][10] Fig. 1. This is an optical effect of the dielectrically expanded skin layer to differentiate from the Debye-sheath plasma mechanism of the electric double layer.

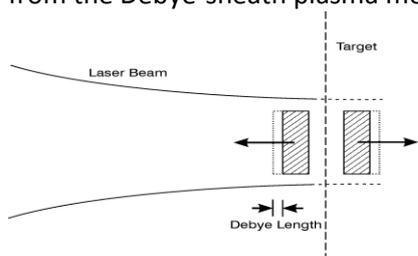

Figure 1. Scheme of skin depth laser interaction where the non-linear force accelerates a plasma block against the laser light and another block towards the target interior. In front of the blocks are electron clouds of the thickness of the effective Debye lengths.

These are the results of plasma-hydrodynamic computations and the ultrahigh acceleration of the plasma block moving against the laser light was seen by a blue Doppler line shift of reflected light. What still needs to be underlined is the fact of the blue shift while numerous experiments showed a red shift. Red shift happened always if the laser interaction was a plasma of lower than critical density confirmed by PIC (particle in cell) computations and numerous measurements. Red shift is a sign of ordinary radiation pressure acceleration RPA as a first easy understanding of this common process. The blue shift is the result – in contrast to the ordinary radiation pressure – to confirm the dielectric explosion of the two plasma blocks, Fig. 1 due to the nonlinear force driven ultrahigh acceleration process. The evaluation of the measurement of Sauerbrey [31] with the blue shift [36] resulted in a dielectric swelling factor of 3.5 that is well known from similar experimental conditions.

The condition of higher densities for explaining the blue shift caused complications for the PIC calculations which were mastered not before [33][34][35]. For a completion or understanding of these conditions, PIC computations were performed for targets with densities close to the critical value where the optical constants are considerably deviating from the vacuum values [33][34].

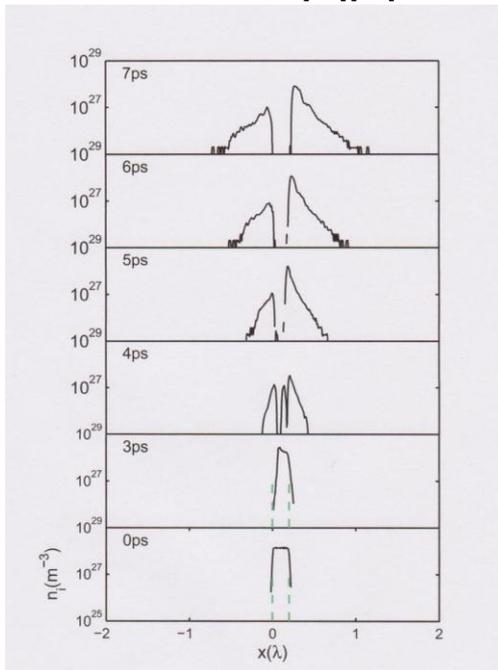

Figure 2a, Spatial distribution of the electron and ion densities under different time after an intense laser pulse is injected upon a solid-density plamsa. The laser intensity $I_o = 10^{15}$ W/cm$^2$, wavelength $\lambda = 0.4\mu m$ and pulse length $\tau = 5ps$. The plasma density $n_o = 2n_c = 1,392 \times 10^{28}$ m$^2$, where $n_c$ is the critical plasma density, and the plasma thickness is $0.2\lambda$. The dashed line marks the initial plasma position.

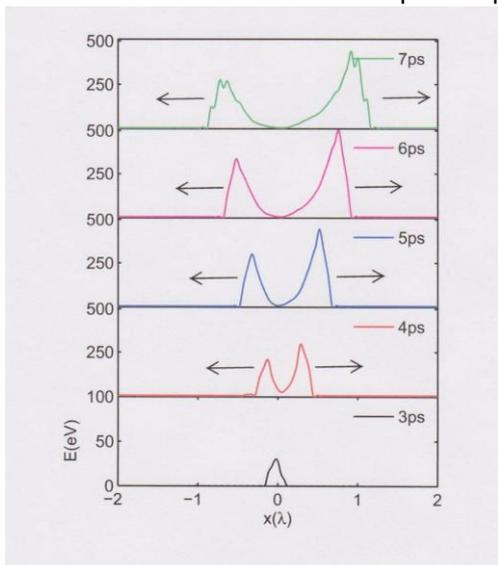

Figure 2b. Results of the electron temperature in the plasma for the cases of Figure 2a.

From Figure 2a, we have clearly observed two neutral plasma blocks (confirmed by printouts for electrons being very close to Fig. 2a) flying in opposite direction from the time t = 3ps when the implosion begins to happen [35]. Fig. 2b shows the corresponding energy evolutions. The predominance of the nonlinear (ponderomotive) force acceleration can be evaluated from the profiles in the figures of the plasma blocks. All the numerical results were carried out with 1D PIC simulation software package VORPAL.

IV. SINGLE LASER BEAM FUSION IGNITION IN REACTOR FOR LASER BORON FUSION

The results with the PIC treatment confirm the preceding hydrodynamic results [7][35] about the side-on direct drive [37] generation of the fusion flame for ignition of solid density fusion, in experimental agreement with the blue Doppler shift [31]. For the step from the plane geometry assumptions to the case of the interaction of a laser beam of given diameter, a radial trapping of a cylindrical fusion fuel is used by magnetic fields. These are in the range above kilotesla and produced by a laser pulse of kilojoule energy and nanosecond duration focussed into one hole between two condenser plates connected by coils [38], see Figure 3. The field of few kilotesla in the coil is strong enough to trap the plasma of the fuel after end-on irradiation of the igniting laser picoseconds laser pulse of more than petawatt power as shown from hydrodynamic computations (see Figures 10.13 to 10.23 in [7])[39]

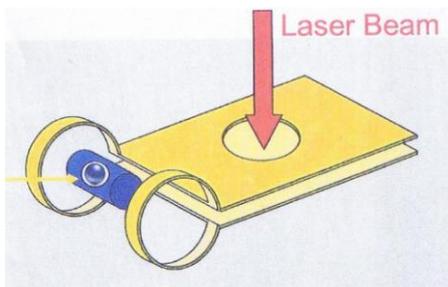

**Figure 3.** Cylindrical solid density HB11 fuel in the coil axis of a magnetic field of few kilotesla produced by the discharge from the condenser plates that is generated by al kilojoule-ns laser pulse [39] irradiated in the hole of the upper plate. For igniting the fusion in the fuel, a laser pulse of ps duration and more than 10 petawatt power is irradiated at the end of the fuel cylinder[19].

This is the step for laser ignition of *fusion by one beam* in difference to the needed radial irradiation for compression of the fuel by the usual scheme with nanosecond pulses or for fast ignition. The design of a fusion reactor is then shown in Fig. 4 where the reaction unit of Fig. 3 is located in the center needing then only one beam for the ignition apart for the nanosecond laser pulse for the generation of the ultrahigh magnetic field. The generated alpha particles of the HB11 reaction destroy the reaction unit at each shot for the pulsating fusion energy production. The generated 2.9 MeV fusion energy in each of alpha particle can be absorbed when entering the spherical wall of the reactor from usual heat exchangers for power generation.

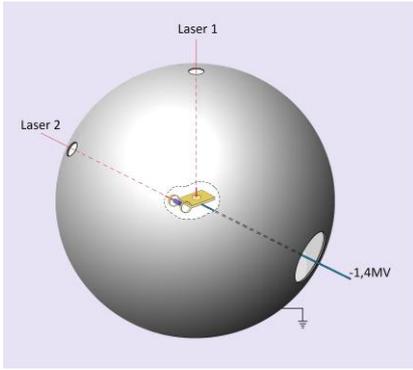

**Figure 4.** Scheme of an economic electric power station for production of boron-fusion, absolutely free from the problem of dangerous nuclear radiation [19] using a reaction unit in the center, Figure 3, for generating the few kilotesla magnetic field by the ns pulse of laser 1 and by a direct drive single ps pulse of laser 2 of more than 10PW power irradiated on the circular end of the fuel.

The advantage of equal energy of the alphas can be used for a direct conversion nearly without heat losses into electricity by letting them move against a static field of up to 1.4 MV, if the reaction units in the center of the wall has such negative voltage against the wall through the support guiding them into the center. The conversion of this dc electric energy into three phase ac is a technique developed for electric power by HVDC (high voltage direct current) transmission [19]. In case that the low density plasma in the sphere causes discharges and would not permit the -1.4 MV charging of the reactor units, then the less efficient option without charging and with heat exchangers can be used.

Due to the fact at comparably low energy of the generated alpha particle and the reaction conditions with sufficiently low temperatures the energy leased in the HB11 reaction [6] are equal energetic of initial 2.9 MeV in difference to more complicate reactions than in Eq. (1) at higher energies. Direct electrostatic conversion of the nuclear fusion energy into electric power with a minimum of losses by heat has been considered by charging the reaction unit at -1.4 Megavolts against the spherical wall of the reactor. The properties of the low density plasma within the sphere and the field with a Faraday cage and related currents and discharges need to be evaluated. In the worst case no charge of the reaction unit against the wall can be used however resulting in a lower efficiency of the power production via thermal energy conversion from the wall.

V. AVALANCHE BORON FUSION IN EXTREME NON-EQUILIBRIUM PLASMA

Up to this point, the fusion of HB11 was considered as a binary reaction as that of DT. In reality, the generation of three alphas, Eq. (1) was always interesting whether subsequent multiplying reactions can increase the energy gain. This is not possible for alphas moving through low temperature non-ionized materials with the well known stopping length in the range of the order of 10μm in solids. But this is different within plasmas of solid state densities of more than several eV temperatures. These conditions of extreme thermal non-equilibrium plasmas [11] permit sufficiently long stopping lengths [7][27]. Measured very high fusion reaction gains [21][22][23] indicated that this is only possible by an avalanche process. The further measurement of more than 1000 times higher fusion gains [22][23] – higher than at comparable conditions of DT fusion were a first prove for the avalanche reactions [19]. The very long stopping lengths of the alphas in plasmas with or without thermal equilibrium are a common knowledge in experiments and simulations [32a].

Due to the fact at comparably low energy of the generated alpha particle and the reaction conditions with sufficiently low temperatures, the energy leased in the HB11 reaction [6] are equal energetic of initial 2.9 MeV in difference to more complicate reactions than in Eq. (1) at higher energies. Direct electrostatic conversion of the nuclear fusion energy into electric power with a minimum of losses by heat has been considered by charging the reaction unit at -1.4 Megavolts against the spherical wall of the reactor. The properties of the low density plasma within the sphere and the field with a Faraday cage and related currents and discharges need to be evaluated. In the worst case no charge of the reaction unit against the wall can be used however resulting in a lower efficiency of the power production via thermal energy conversion from the wall.

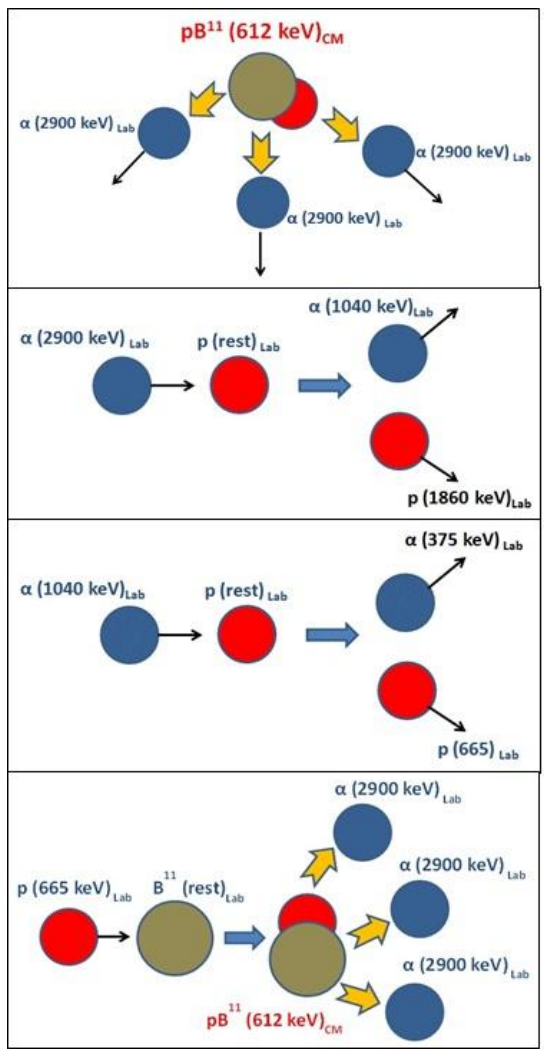

Fig. 5. A schematic overview of the avalanche process.

The specific evaluation of elastic collisions of the generated alphas with protons and boron nuclei [41] documented how the hydrogen nuclei receive an energy within a wide range around 600 keV energy, Figure 5, for reacting with the $^{11}$B nuclei at nearly ten times higher fusion cross sections compared with all known other fusion reactions [8], to produce each three alphas etc. for the avalanche. The measurements [22] with the nuclei for the energetic HB11 reactions on the background of less than few ten eV background plasma, could be theoretical reproduced in details [41]. This shows the need to explore this kind of non-ideal [42] and non-neutral [43] plasmas. The earlier estimations of the anticipated avalanche reactions [40] was then fully proved for use [39]. Under simplified assumptions [39], the reaction of 12 mg boron fuel can produce one GJ = 277 kWh or more fusion energy, ignited in controlled way by the one single ps irradiated laser beams in the reactor of Figure 4. The easy operation with one beam ignition should then permit a reactor with one shot per second [44] and sufficiently fast localisation of the reaction unit using presently available technology for low cost power generation [37][45]. The now presented results show an increase of the HB11 fusion gains by more than nine orders of magnitudes above the classical value [41].

Fusion energy for block-ignition by ultrahigh power ps laser pulses. News Release Austr. Physics Congress December 2014.